\documentclass[fleqn,12pt,twoside]{article}
\usepackage[headings]{espcrc1}

\readRCS
$Id: espcrc1.tex,v 1.2 2004/02/24 11:22:11 spepping Exp $
\usepackage{graphicx}
\usepackage[figuresright]{rotating}

\newcommand{\AmS}{{\protect\the\textfont2
  A\kern-.1667em\lower.5ex\hbox{M}\kern-.125emS}}

\title{Is there a "Sonic Boom" in the Little Bang at RHIC ?}

\author{ Roy. A. Lacey\address[SbChem]{Chemistry Department, 
        Stony Brook University, \\ 
        Stony Brook, New York, USA}
      }
       
\runtitle{Is there a "Sonic Boom" in the Little Bang at RHIC ?}
\runauthor{Roy A. Lacey}

\begin{document}

% typeset front matter
\maketitle

\begin{abstract}
	The use of elliptic flow and correlation measurements as 
constraints to establish the transport properties of hot and dense QCD  
matter is discussed. Measured Two- and three-particle correlation functions 
give initial hints for a ``sonic boom". Elliptic flow measurements give the 
estimates $c_s \sim 0.35$ and $\eta/s \sim 0.1$ for the sound speed and 
viscosity to entropy ratio.
\end{abstract}

\section{Introduction }

The Big Bang Model is a widely accepted theory for the origin and evolution 
of our universe. It postulates that $\sim 12-14$ billion years ago,  
a hot and dense plasma of quarks and gluons (QGP) was produced a few micro seconds 
after the ``big bang". A pervading uniform glow of cold cosmic microwave 
background provides an unequivocal signature for the subsequent 
expansion from this hot and dense state to the vast and significantly cooler 
cosmos we currently inhabit \cite{Peebles91}.
 
	Recent experiments at Brookhaven's Relativistic Heavy Ion Collider (RHIC), 
give evidence for the creation of locally equilibrated hot and dense QCD matter 
in a ``little bang" initiated in relativistic Au+Au collisions \cite{Adcox:2004mh}. 
A short time after this bang ($\sim 1fm/c$), the energy density is estimated to 
be  $\simeq 5.4$~GeV/fm$^3$ \cite{Adcox:2004mh} 
-- a value significantly larger than the $\sim 1$~GeV/fm$^3$ required for the 
transition from low-temperature hadronic matter to the high-temperature QGP phase 
of QCD \cite{Karsch:2001vs}. The subsequent expansion and ultimate hadronization 
of this high energy density matter, results in the emission of particles which 
signal its thermodynamic and dynamical properties. 

	An important challenge is to develop robust experimental constraints for these 
properties. This contribution discusses the role of both elliptic flow and 
hot QCD ``sonic boom" measurements as constraints.

\section{Sonic Booms}

\begin{figure}[t]
\begin{minipage}[t]{80mm}
%\framebox[79mm]{\rule[-26mm]{0mm}{52mm}}
\includegraphics[width=0.8\linewidth]{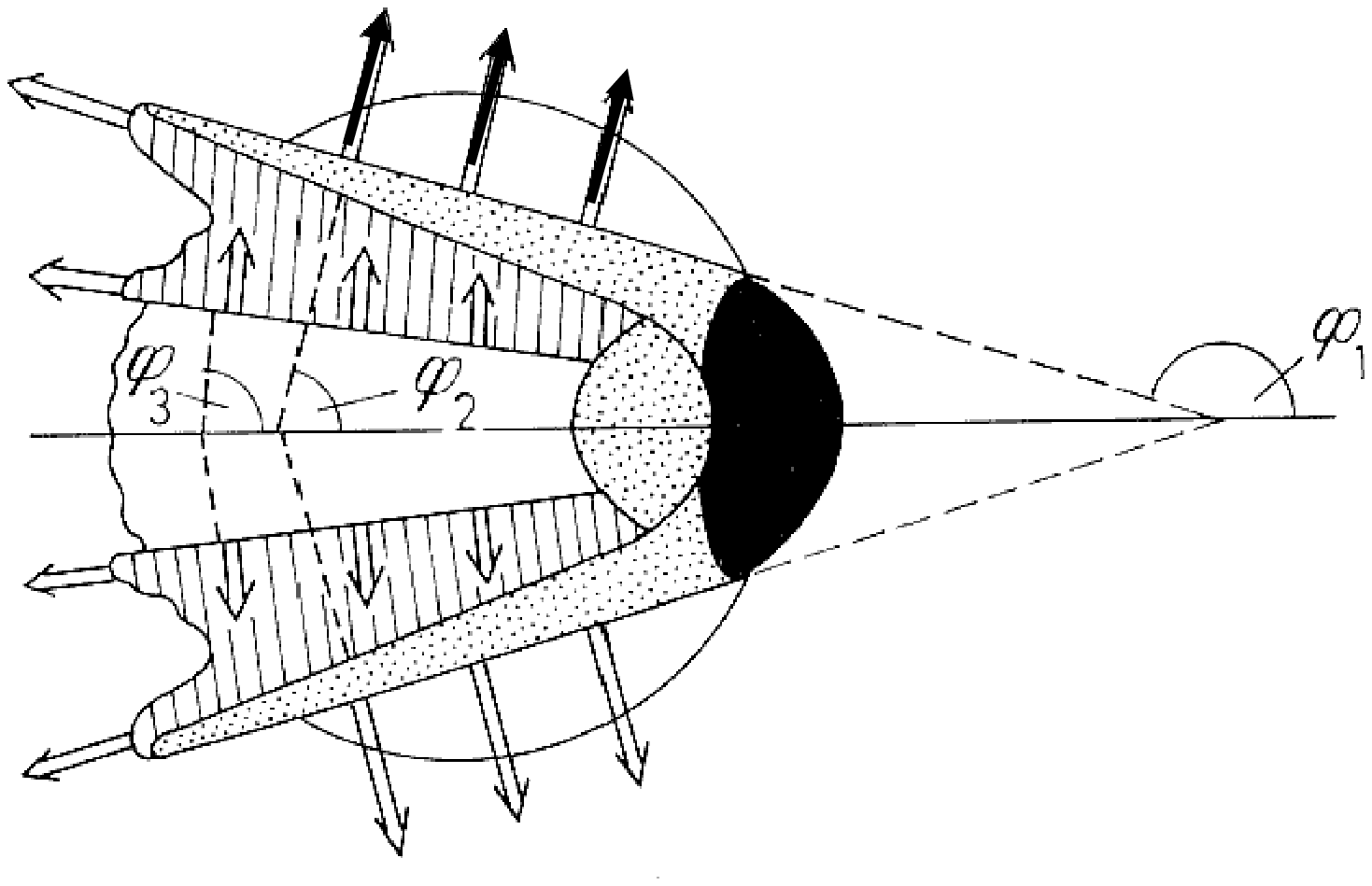}
\vskip -0.75cm
\caption{Schematic illustration of the development of a Mach cone 
resulting from the bombardment of a heavy target with a relativistic 
light projectile \cite{early_conical_flow}. 
}
\label{fig:shock}
\end{minipage}
\hspace{\fill}
\begin{minipage}[t]{75mm}
%\framebox[74mm]{\rule[-26mm]{0mm}{52mm}}
\includegraphics[width=0.70\linewidth]{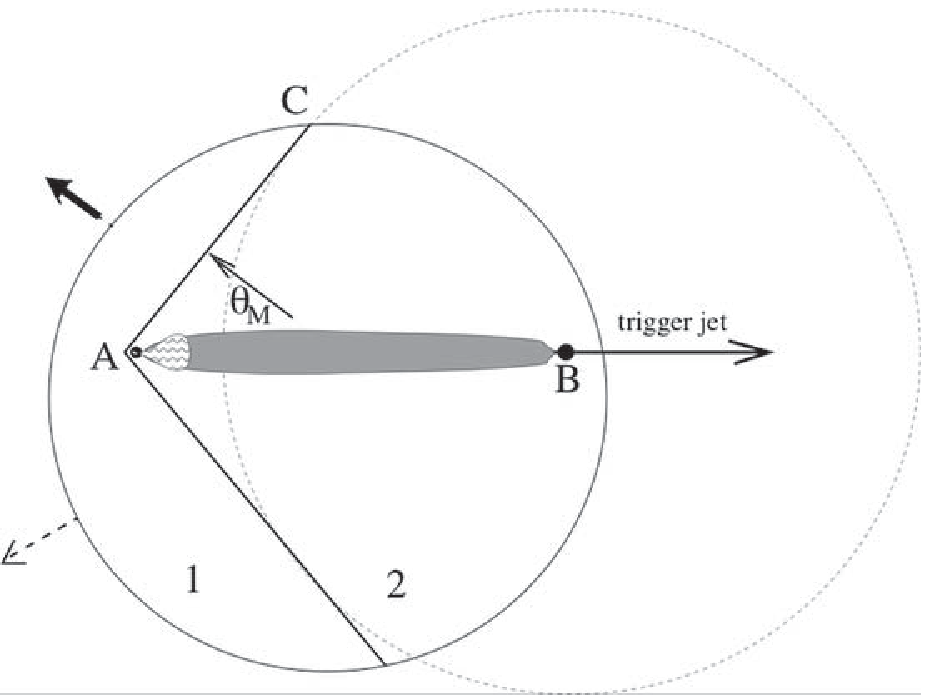}
\vskip -0.75cm
\caption{Schematic diagram of the development of a Mach cone resulting from 
the strong interaction between a jet and a high energy density medium \cite{mach_cones}.
%The arrows which point to the left indicate the direction of flow normal to the shock front. 
}
\label{fig:mach_cone}
\end{minipage}
\end{figure}

	Objects moving at supersonic speeds in a medium create a wake behind the shock
front they create. The familiar sonic boom from supersonic jets is the audible component 
of such a Mach cone produced in air. Similar Mach cones were initially predicted to occur 
in cold nuclear matter via bombardment of a heavy target nucleus with a light relativistic 
projectile \cite{early_conical_flow}. A schematic illustration for the generation of 
such a shock front is shown in Fig. \ref{fig:shock}. Matter flow is normal to the shock 
front which subtends a Mach angle given by the ratio $v/c_s$, where $v$ and $c_s$ are the 
projectile and sound velocities.

	An extensive search at the BEVALAC concluded that cold nuclear matter was too 
dilute and dissipative to sustain the propagation of such a shock front. 

\subsection{Sonic Booms in Hot and Dense QCD Matter}

	A confluence of recent RHIC measurements have generated much interest in the search 
for Mach cones in high energy density QCD matter \cite{mach_cones}: 
(i) Hot and dense QCD matter is produced well above the temperature 
required for QGP production. 
(ii) Elliptic flow measurements validate the 
predictions of perfect fluid hydrodynamics \cite{Issah:2006qn,Adare:2006ti} and do 
not show the expected suppression of flow due to the shear viscosity predicted 
by weak-coupling transport calculations. In fact, only a rather small value for the 
viscosity to entropy ratio $\eta/s \sim 0.1$, can be accommodated by the data 
(see estimate below and Refs.~\cite{Teaney:2003kp} and \cite{Gavin:2006xd}). 
This has been interpreted as evidence that the QGP created in the early phase 
of RHIC collisions is more strongly coupled than expected \cite{Shuryak:2004cy}. 
(iii) Jet suppression measurements \cite{jet_quenching} indicate that the hot 
and dense QCD matter is almost opaque to high energy partons which deposit a 
substantial fraction of their energy into the matter. 

	Thus, the expectation is that the jet energy dumped into the strongly 
coupled plasma could survive as a coherent radiation of sound waves leading 
to a Mach cone (sonic boom) with Mach angle $\cos\theta_M = v_j/c_s$ as 
illustrated in Fig.~\ref{fig:mach_cone} \cite{mach_cones}. Here, $v_j$ is the 
jet velocity and the direction of matter flow is normal to the shock front.
An important point to be made here is that the detection of a sonic boom naturally
leads to important constraints for the viscosity and sound speed of hot QCD matter. 

	Two- and three-particle correlation function measurements provide 
an excellent probe for the sonic boom \cite{3pc_ajit}.
\begin{figure}[t]
\begin{minipage}[t]{80mm}
%\framebox[79mm]{\rule[-26mm]{0mm}{52mm}}
%\vskip -0.5cm
\includegraphics[width=1.\linewidth]{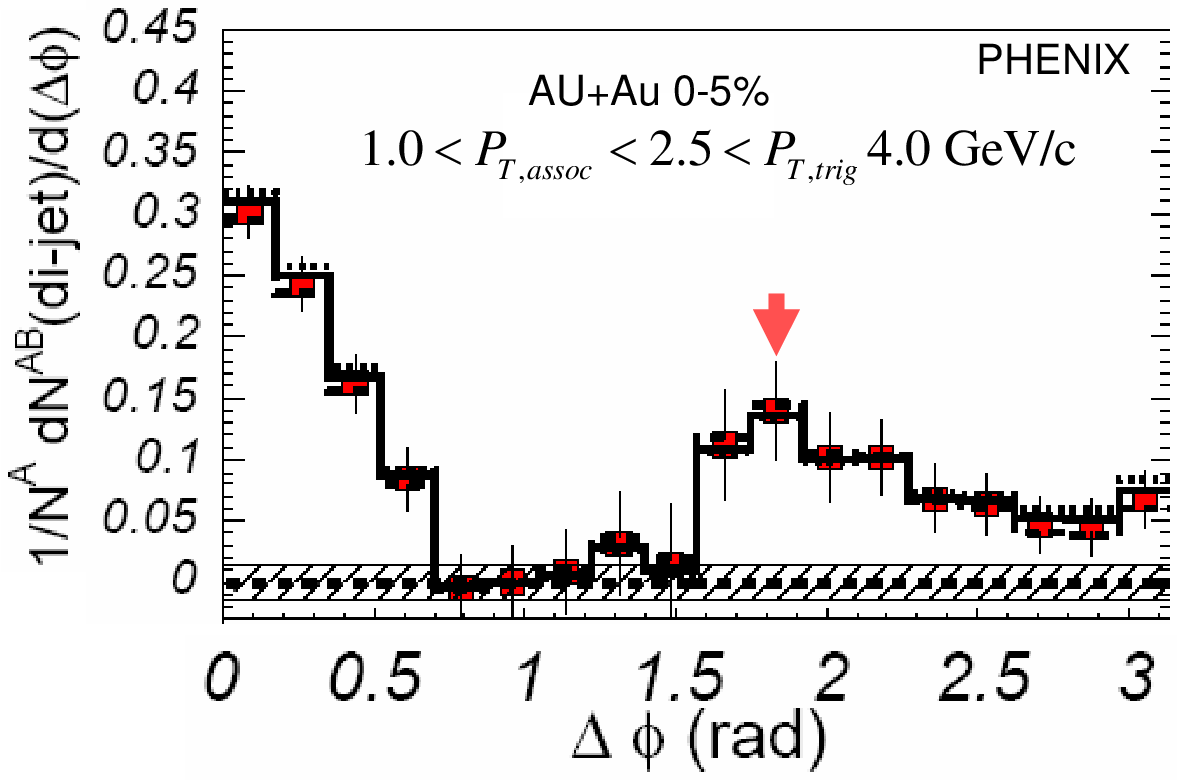}
\vskip -1.00cm
\caption{(Color online) Background subtracted $\Delta\phi$ 
correlation function for central Au+Au collisions obtained by the PHENIX experiment. 
}
\label{fig:2pc_phenix}
\end{minipage}
\hspace{\fill}
\begin{minipage}[t]{75mm}
%\framebox[74mm]{\rule[-26mm]{0mm}{52mm}}
%\vskip -0.5cm
\includegraphics[width=1.2\linewidth]{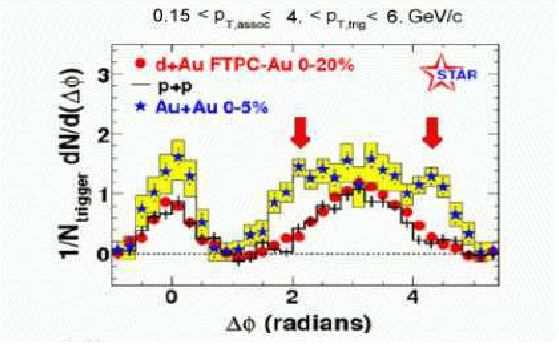}
\vskip -1.00cm
\caption{(Color online) Background subtracted $\Delta\phi$ correlation 
function for central Au+Au, d+Au and p+p collisions obtained by the STAR experiment.  
}
\label{fig:2pc_star} 
\end{minipage}
%\vskip -0.3cm
\end{figure}
Background subtracted $\Delta\phi$ correlation functions from central 
Au+Au collisions are shown in Figs. \ref{fig:2pc_phenix} and \ref{fig:2pc_star} 
for trigger and associated particle $p_T$s as indicated. In contrast to 
p+p and d+Au collisions which show an away-side jet peak at 
$\Delta\phi = 180^o$ (cf. Fig.~\ref{fig:2pc_star}), the correlation functions
for central Au+Au collisions show a broad minimum (cf. Fig.~\ref{fig:2pc_phenix})
at this angle. An apparent maximum in the distribution for the away-side jet 
corresponding to an estimated Mach angle (see red arrows) has been interpreted 
as initial evidence for a sonic boom in hot QCD matter \cite{mach_cones}.

	An issue which complicates the Mach cone interpretation of the Au+Au results 
in Figs. \ref{fig:2pc_phenix} and \ref{fig:2pc_star} is the possibility that the 
away-side jet might undergo large deflections due to interactions between the 
jet and the expanding system \cite{Armesto:2004vz}. Such deflections can 
give rise to a $\Delta\phi$ correlation function similar to the one expected 
from a Mach cone. To resolve this ambiguity, state-of-the-art three-particle 
correlation function techniques have been developed \cite{3pc_ajit}. 

\begin{figure}[t]
%\begin{minipage}[t]{80mm}
%\framebox[79mm]{\rule[-26mm]{0mm}{52mm}}
%\vskip -0.3cm
\includegraphics[width=1.\linewidth]{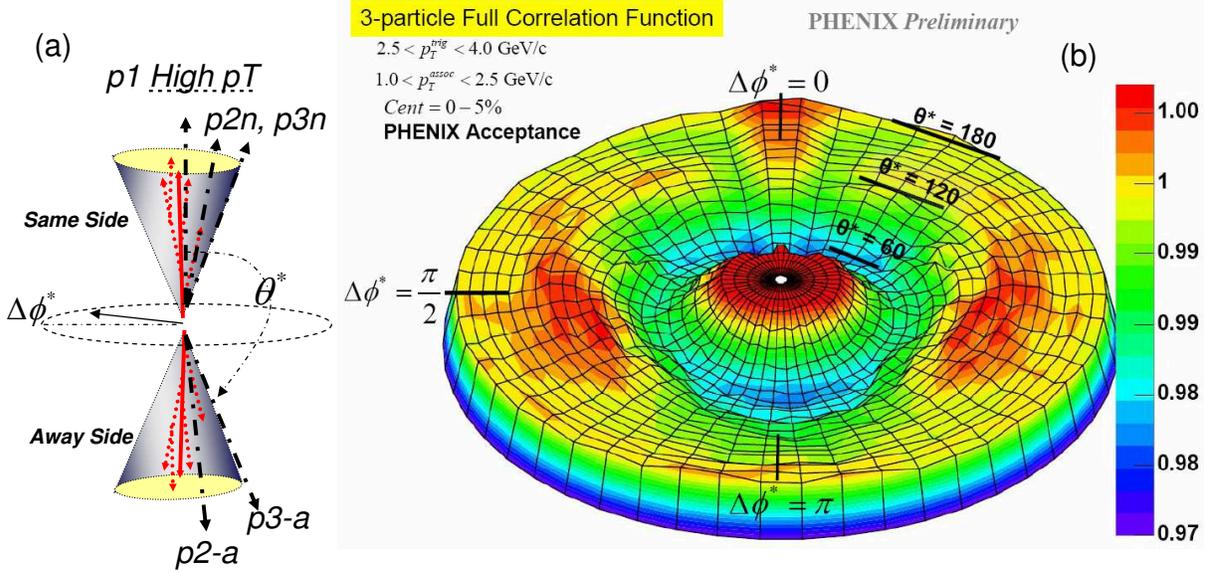}
\vskip -0.6cm
\caption{(Color online)(a) Schematic diagram of the coordinate frame used for three-particle 
correlation functions. The leading (high $p_T$ partcle) is assumed to be the $z$ axis. 
The polar coordinates $\theta^*,\Delta\phi^*$ of the associated particles are indicated.
(b) Full $\theta^*,\Delta\phi^*$ three particle correlation surface for charged hadrons 
detected in central (0-5\%) Au+Au collisions within the PHENIX acceptance. 
}
\label{3pc_cor_0-5}
\end{figure}

	Figure \ref{3pc_cor_0-5}b shows a full three-particle correlation 
function (no background subtraction) obtained by combining a high $p_T$ particle 
with two associated low $p_T$ particles in a frame which approximates 
the actual jet frame. This frame of reference is illustrated in 
Fig. \ref{3pc_cor_0-5}a. The high $p_T$ trigger particle ($p1$) is coincident 
with the z-axis. The polar angle $\theta^*$, and the azimuthal angle difference 
$\Delta\phi^*$ for the associated low $p_T$ particles ($p2n$ and $p3n$ for 
same (near) side jet; $p2$-a and $p3$-a for the away-side jet) are used as  
correlation variables. In this polar ($\theta^*,\Delta\phi^*$) representation, 
the near-side jet correlations are expected at the center of the  
correlation surface; after background subtraction, a $\Delta\phi^*$ ring 
at $\theta^* = v_j/c_s$ (ie. the Mach angle) would signal the correlations 
from an away-side Mach cone. 

	Figure \ref{3pc_cor_0-5}b show sizable correlations in $\Delta\phi^*$ for 
$\theta^* \sim 120^o$, albeit with acceptance losses especially in the region 
about ($\Delta\phi^* \sim 180^0$). These correlations are tantalizingly 
suggestive. However, the answer to the question as to whether or not they 
are due to a Mach cone, must await the results of accurate background 
subtraction. Suffice to say, initial indications are quite 
encouraging\footnote{In a recent presentation at the Hard Probes meeting, 
the STAR collaboration has also indicated a significant three-particle correlation 
yield after background subtraction.}.

\section{Transport Properties from $v_2$ Measurements}
\begin{figure}[t]
\begin{minipage}[t]{80mm}
%\framebox[79mm]{\rule[-26mm]{0mm}{52mm}}
%\vskip -0.3cm
\includegraphics[width=0.8\linewidth]{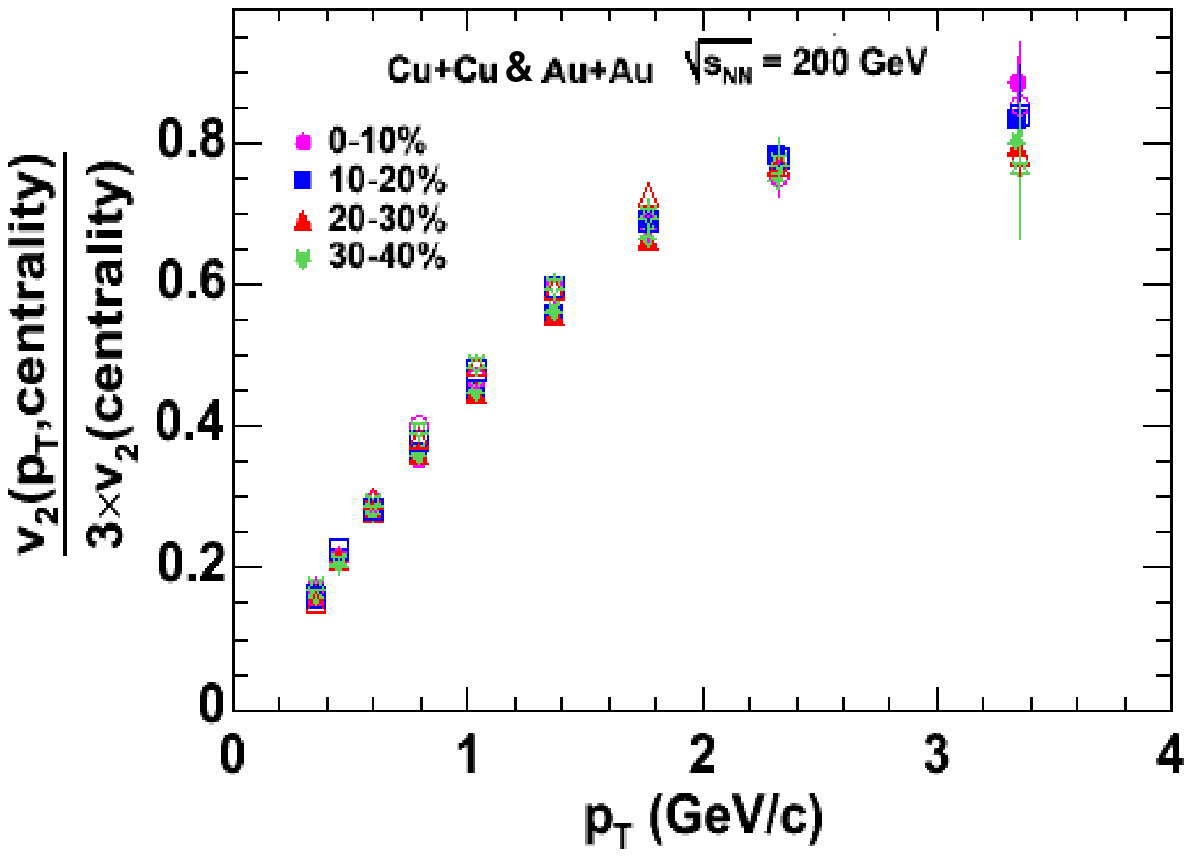}
\vskip -0.5cm
\caption{ (Color online) $v_2(centrality, p_T)$ divided by 3 times the $p_T$-integrated 
value $v_2(centrality)$, for Au+Au and Cu+Cu  collisions.
}
\label{ecc_scaling_Au-Cu}
\end{minipage}
\hspace{\fill}
\begin{minipage}[t]{75mm}
%\framebox[74mm]{\rule[-26mm]{0mm}{52mm}}
%\vskip -0.3cm
\includegraphics[width=0.7\linewidth]{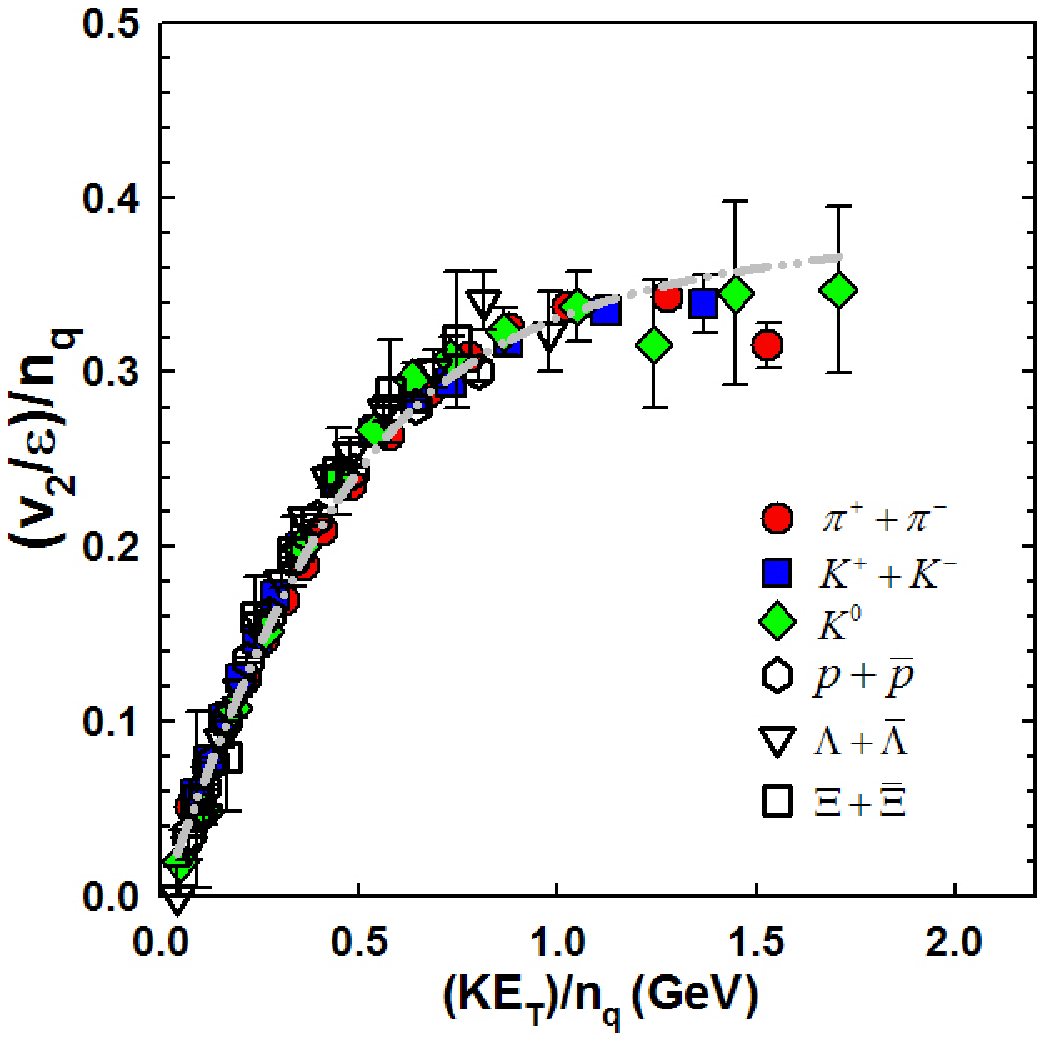}
\vskip -0.5cm
\caption{(Color online) $v_2/\varepsilon n_q$ vs $KE_T/n_q$ for several identified 
particle species for minimum bias Au+Au collisions. 
}
\label{scaled_v2_fit}
\end{minipage}
%\vskip -0.3cm
\end{figure}

As mentioned earlier, elliptic flow measurements \cite{Issah:2006qn,Adare:2006ti} validate the 
predictions of perfect fluid hydrodynamics for the scaling of the elliptic 
flow coefficient $v_2$ with eccentricity $\varepsilon$, system size and 
transverse kinetic energy $KE_T$ \cite{Csanad:2005gv,Bhalerao:2005mm,Csanad:2006sp}; 
they also indicate the predictions of quark number ($n_q$) scaling \cite{Voloshin:2002wa},
suggesting that quark-like degrees of freedom are pertinent when elliptic 
flow develops. 

	Figure \ref{ecc_scaling_Au-Cu} show scaled $v_2$ values for Cu+Cu and Au+Au 
collisions; they are clearly independent of the colliding system size and show 
essentially perfect $\varepsilon$ scaling for a broad range of centralities.
Such scalings are in accord with the scale invariance of perfect fluid 
hydrodynamics \cite{Lacey:2005qq,Bhalerao:2005mm} and can be used to 
constrain the sound speed  \cite{Issah:2006qn,Adare:2006ti}.

	Figure \ref{scaled_v2_fit} shows a plot of  $v_2/n_q\varepsilon$ vs $KE_T/n_q$; 
it demonstrates that the relatively ``complicated" dependence of $v_2$ on 
centrality, transverse momentum, particle type and quark number can be 
scaled to a single function \cite{Issah:2006qn,Adare:2006ti}. 
The dashed-dotted curve indicates a fit to these data.

\subsection{Estimating $\eta/s$ }

	The viscosity to entropy ratio can be expressed as:
\[
\; \; \; \eta/s \sim T \lambda_{f} c_s, 
\]
where $T$ is the 
temperature, $\lambda_{f}$ is the mean free path and $c_s$ 
is the sound speed in the matter. The temperature $T=165\pm 3$~MeV is constrained 
via a fit to the data in Fig.~\ref{scaled_v2_fit} with the fit function 
$I_1(w)/I_0(w)$, where $w=KE_T/2T$ and $I_1(w)$ and $I_0(w)$ 
are Bessel functions \cite{Csanad:2006sp}. For $c_s$, the estimate 
$c_s = 0.35 \pm 0.05$, given in Refs.~\cite{Issah:2006qn} and \cite{Adare:2006ti}
is used. An estimate $\lambda_f = 0.3 \pm 0.03 fm$, is obtained from the 
on-shell transport model simulations of Xu and Greiner \cite{Xu:2004mz}. 
This parton cascade model includes pQCD $2 \leftrightarrow 2$ and 
$2 \leftrightarrow 3$ scatterings. 

	With these values for $\lambda_{f}$, $T$ and $c_s$ one obtains the 
estimate $\eta/s \sim 0.09 \pm 0.015$ which is in good agreement with the 
estimates of Teaney and Gavin \cite{Teaney:2003kp,Gavin:2006xd}.
It is also close to the lower bound of $\eta/s = 1/4\pi$, reached in the 
strong coupling limit of certain gauge theories~\cite{Kovtun:2004de}.

\section{Summary}

		The development of robust experimental constraints for the 
thermodynamic and transport properties of the strongly coupled plasma 
believed to be produced in Au+Au collisions at RHIC, constitutes 
a major current challenge. Elliptic flow measurements provide an initial 
set of constraints for the estimates $c_s \sim 0.35$ and $\eta/s \sim 0.1$. 
Quantitative confirmation of an initial hint for a hot QCD ``sonic boom" 
from two- and three-particle correlation functions, will undoubtedly 
provide invaluable additional constraints.

\end{document}